\documentclass[a4paper,11pt]{article}
\pdfoutput=1 
\usepackage{jheppub} 
\usepackage{verbatim}

\usepackage{graphicx}	
\usepackage{dcolumn}	
\usepackage{bm}			
\usepackage{hyperref}
\usepackage{amsmath}
\usepackage{verbatim}
\usepackage{slashed}
\usepackage{hyperref}
\usepackage{mathtools}
\usepackage{tabularx}

\title{\boldmath Explaining the $R_{K^{(*)}}$ anomalies and the CDF $M_W$ in Flavorful Top Seesaw Models with Gauged $U(1)_{L(-R)}$}

\author{Yi Chung}

\affiliation{
Max-Planck-Institut für Kernphysik, Saupfercheckweg 1, 69117 Heidelberg, Germany
}

\emailAdd{yi.chung@mpi-hd.mpg.de}

\abstract{A new heavy $Z'$ vector boson provides a possible explanation for the neutral current B anomalies and points to the scale where we expect the solution to the hierarchy problem to appear. In this paper, we explore a modified Top Seesaw model based on the global symmetry breaking of $U(3)_L$ to $U(2)_L$. The symmetry is partially gauged such that the breaking introduces not only a composite Higgs doublet but also a TeV-scale $Z'$ boson. The $U(3)_L/U(2)_L$ Top Seesaw model predicts a light Higgs boson as well as a large top Yukawa coupling. Additional $U(1)'$ symmetry is gauged such that a flavorful TeV-scale $Z'$ boson is introduced to address the B anomalies. The existence of a heavy vector-like top quark and $Z-Z'$ mixing breaks the custodial symmetry, which predicts a heavier $W$ mass as observed by the CDF collaboration. Two possible candidates for the $U(1)'$ symmetry corresponding to the $Z'$ boson, $U(1)_L$ and $U(1)_{L-R}$, are discussed in detail. The parameter space for the models to address both anomalies is presented. The direct searches of the $Z'$ boson from dilepton channels are also studied. Based on the assumptions, a TeV-scale $Z'$ boson is still viable and can be probed in the near future.}

\begin{document}
\maketitle
\flushbottom


\section{Introduction}


The Standard Model (SM) of particle physics successfully describes all known elementary particles and their interactions. With the discovery of light Higgs bosons in 2012 \cite{Chatrchyan:2012xdj, Aad:2012tfa}, the last missing piece of the SM seemed to be filled. However, the SM does not address the UV-sensitive nature of scalar bosons. The naturalness principle predicts some new physics not too far from the electroweak scale. Many models aiming to solve the hierarchy problem have introduced new particles with various properties and phenomenology around the TeV scale. But yet, none of them have been observed, which makes the hierarchy problem more challenging and profound.


Despite the lack of evidence for new particles from direct searches by ATLAS and CMS collaboration, a set of interesting discrepancies emerged from the precise measurements of B-meson semileptonic decays by LHCb collaboration. For example, the measurements of lepton flavor universality (LFU) ratio $R(K)\equiv BR(B\to K\mu^+\mu^-)/BR(B\to Ke^+e^-)$ and $R(K^*)$ show growing hints of beyond the Standard Model (BSM) physics \cite{LHCb:2014vgu, LHCb:2019hip, LHCb:2021trn, LHCb:2017avl} \footnote{The hints are completely gone based on the latest measurement \cite{LHCb:2022zom}.}. Other related measurements include angular observables of the related decay \cite{LHCb:2013ghj, LHCb:2015svh, LHCb:2020lmf}. Each anomaly is not statistically significant enough to reach the discovery level, but the combined analysis shows a consistent deviation from the SM prediction \cite{Altmannshofer:2021qrr, Cornella:2021sby, Geng:2021nhg, Alok:2019ufo, Alguero:2021anc, Carvunis:2021jga, Hurth:2021nsi}. They are known as neutral current B anomalies (NCBAs). The global fit results point to the additional contribution to the operators
\begin{equation}
\Delta\mathcal{L}_{\text{eff}}=\frac{4G_F}{\sqrt{2}}V_{tb}V_{ts}^*\frac{e^2}{16\pi^2}
~C_{9(10)}(\bar{s}\gamma_\mu P_Lb)(\bar{\mu}\gamma^\mu(\gamma^5)\mu)
\quad\text{with}\quad |C_{9(10)}|\sim 1~,
\end{equation}
which implies a generic scale of new physics $\Lambda_{\text{NP}}\sim 35$ TeV. However, if the couplings receive some suppressions such as small rotation angles, the scale can be brought down to a few TeV, which is where we now expect a solution to the hierarchy problem!

Among all kinds of possible explanations for the NCBAs, the two tree-level mediators, Leptoquarks and $Z'$ bosons, are the most promising solutions. Leptoquarks, however, are not the common ingredient in the models for the hierarchy problem. Some efforts have been made to connect leptoquarks to the electroweak scale \cite{Gripaios:2014tna, Barbieri:2016las, Blanke:2018sro, Marzocca:2018wcf, Fuentes-Martin:2020bnh}, but some extensions to connect quarks and leptons are required. On the other hand, a $Z'$ boson from a broken $U(1)'$ gauge symmetry is common in many BSM models. Therefore, we look for an economical model which predicts a Higgs doublet together with a TeV-scale $Z'$ boson as the lowest states in the spectrum. We find that such a spectrum can naturally be realized in specific Composite Higgs Models (CHM).


In a CHM, the Higgs doublet is the pseudo-Nambu-Goldstone boson (pNGB) of a spontaneously broken global symmetry of the underlying strong dynamics~\cite{Kaplan:1983fs, Kaplan:1983sm}. Analogous to the chiral symmetry breaking in quantum chromodynamics (QCD), where light scalar particles, pions, are naturally introduced, we can build models with light Higgs bosons in a similar way. In a CHM, an approximate global symmetry $G$ is spontaneously broken down to a subgroup $H$ by some strong dynamics at a symmetry breaking scale $f$. The heavy resonances of the strong dynamics are expected to be around the compositeness scale $\sim 4\pi f$ generically. The pNGBs of the symmetry breaking, on the other hand, can naturally be light with masses $<f$ as they are protected by the shift symmetry.

The most well-studied CHM is Minimal Composite Higgs Models~\cite{Agashe:2004rs} with the coset $SO(5)/SO(4)$, which contains exactly four pNGBs. However, to introduce an additional $Z'$ boson in the spectrum, the candidate should be CHMs with an additional Goldstone boson ( i.e., five pNGBs), where the corresponding $U(1)'$ symmetry is gauged. There are only two candidate cosets - $SU(4)/Sp(4)\cong SO(6)/SO(5)$ and $SU(3)\times U(1)/SU(2)\times U(1)$. They both introduce five pNGBs as a complex doublet plus a real singlet under the SM gauge group. To understand the properties of the $U(1)'$ symmetry to be gauged, the UV completion for the CHMs is also required.


Starting with the $SU(4)/Sp(4)$ CHM, the coset is the minimal coset which can be realized with fundamental fermionic matter theory, known as fundamental composite Higgs model (FCHM) \cite{Barnard:2013zea, Ferretti:2013kya, Cacciapaglia:2014uja, Cacciapaglia:2020kgq}. In this type of CHMs, hyperfermions $\psi$ are introduced as the representation of hypercolor (HC) group $G_{HC}$. Once the HC group becomes strongly coupled, hyperfermions form a condensate, which breaks the global symmetry $G$ down to the subgroup $H$. In the $SU(4)/Sp(4)$ FCHM \cite{Katz:2005au, Gripaios:2009pe, Galloway:2010bp}, the broken $U(1)'$ symmetry within the $SU(4)$ global symmetry turns out to be the baryon number of hyperfermions, i.e., hyperbayron number (HB). However, it is anomalous under $SU(2)_L$ and can not be gauged directly. One way to make it anomaly-free is to have the SM fermions charged under the new $U(1)'$ symmetry. The resulting symmetry becomes $U(1)_{SM-HB}$, which is like a hyperversion of anomaly-free $U(1)_{B-L}$ symmetry. It was shown in \cite{Chung:2021ekz, Chung:2021xhd} that if only the third generation SM fermions are charged and the symmetry becomes $U(1)_{SM_3-HB}$, the corresponding $Z'$ boson can provide a good explanation to the NCBAs.


In this paper, we will focus on the other coset, the $SU(3)\times U(1)/SU(2)\times U(1)$ CHM. Unlike FCHMs, this breaking pattern can not be realized with fundamental gauge dynamics featuring only fermionic matter fields. However, a similar structure based on $U(3)_L/U(2)_L$ is discussed in CHMs with top seesaw mechanism \cite{Dobrescu:1997nm, Chivukula:1998wd, Cheng:2013qwa}. The model requires an additional vector-like top quark, which together with the third generation quarks $q_L$ and $t_R$ are charged under some non-confining strong interaction. The Higgs boson is introduced as the bound state of a top quark and a vector-like top quark, which naturally has a large top Yukawa coupling. At the same time, it is also a pNGB of broken global symmetry $U(3)_L/U(2)_L$, so it is naturally light. We will focus on this economical setup\footnote{The version with custodial symmetry was also presented by the same author \cite{Cheng:2014dwa} but required additional exotic vector-like quarks.} and only modify it with an additional gauged $U(1)'$ symmetry, which can lead to a flavorful $Z'$ boson to explain the neutral current B anomalies.


Another issue of the coset $U(3)_L/U(2)_L$ is the lack of protection of custodial symmetry. However, the $W$ boson measurement by the CDF collaboration \cite{CDF:2022hxs} implies that instead of fine protection, we might need to break it. It was pointed out in many following analyses \cite{Lu:2022bgw, Strumia:2022qkt, deBlas:2022hdk, Bagnaschi:2022whn, Asadi:2022xiy} that an additional contribution on the $T$ parameter is required. Several attempts to address the $W$ mass anomaly together with neutral current B anomalies based on $Z'$ solution have been studied \cite{Alguero:2022est, Li:2022gwc, Allanach:2022bik}. Works on explaining the $M_W$ in CHMs have also been done \cite{Cacciapaglia:2022xih, Appelquist:2022qgl, Frandsen:2022xsz, Cai:2022zqu}. In our model, the vector-like top quark and new $Z'$ boson both violate the custodial symmetry and can introduce sufficient $\Delta T$ with the suitable coupling and mass. The electroweak precision test (EWPT), which used to be the strongest constraint on the model, now becomes the measurement to decide the preferred parameter space.


To generate the CDF $W$ mass, specific values of the breaking scale $f$ of the Top Seesaw models are preferred. On the other hand, to explain the neutral current B anomalies without violating other flavor constraints, the $Z'$ scale $f'$ is also bounded. We can then compare the scale to determine the suitable parameter space for each $U(1)'$ symmetry we introduce. We find that there is still viable parameter space for $U(1)_L$ (and $U(1)_{L-R}$ if the NCBAs are $2\sigma$ smaller) to explain both anomalies at once. Based on the different $U(1)'$ symmetry and parameter spaces, we can further calculate the cross section for different decay channels. Compared with the current constraints from direct searches, we show that a $Z'$ boson of $U(1)_L$ with $M_{Z'}>1.25$ TeV is still viable. Even lighter $Z'$ bosons are possible for both $U(1)_L$ and $U(1)_{L-R}$ if NCBAs are $2\sigma$ smaller.


This paper is organized as follows. In section~\ref{sec:Model}, we discuss the properties of $U(3)_L/U(2)_L$ Top Seesaw model. The scalar potential and vacuum structure are presented. The top seesaw mechanism is also shown. Next, we discuss the $U(1)'$ gauge symmetry which can accommodate within the global symmetry $U(3)_L\times U(2)_R$ in section~\ref{sec:U(1)'}. Two candidates, $U(1)_{L}$ and $U(1)_{L-R}$ are chosen to study in this paper. The lack of custodial symmetry in this model leads to a strong constraint from electroweak precision tests but also provides a chance to explain the CDF $W$ mass. The preferred parameter space is calculated in section~\ref{sec:EW}. The flavor physics, including the B anomalies and other experimental constraints, are discussed in section~\ref{sec:Flavor}. The analysis combining the requirements from the two anomalies is shown in section~\ref{sec:Combine}. Next, we turn to the direct searches in section~\ref{sec:Collider}, where we focus on the phenomenology from new $Z'$ bosons. Section~\ref{sec:Conclusion} contains our conclusions.


\section{The $U(3)_L/U(2)_L$ Top Seesaw model}\label{sec:Model}

In Top Seesaw models \cite{Dobrescu:1997nm, Chivukula:1998wd}, an additional $SU(2)_L$-singlet vector-like quark $T$ with the same electric charge as the top quark is introduced. In the original models, a strong interaction is required to generate the $\bar{T_R}\,q_L $ bound state behaving like the SM Higgs doublet. Such a strong interaction needs to be non-confining, otherwise, the top quark would be screened. To introduce the composite Higgs as pNGB \cite{Cheng:2013qwa}, the strong interaction should further involve both the right-handed top quark $t_R$ and left-handed vector-like quark $T_L$, which has the global symmetry to $U(3)_L\times U(2)_R$. We consider a strong interaction mediated by massive gauge bosons, which can arise from an asymptotically-free gauge symmetry broken around the scale $\Lambda$. At the energy below $\Lambda$, the broken gauge interaction will lead to four-fermion interaction with $1/\Lambda^2$ suppression. However, if the coupling is strong enough, the fermion condensate can still be formed, which can lead to the global symmetry breaking $U(3)_L \to U(2)_L$\footnote{In general, the global symmetry breaking should be $U(3)_L\times U(2)_R \to U(2)_L\times U(1)_R$ with total 8 pNGBs. However, to realize the top seesaw mechanism, the $U(2)_R$ needs to be broken explicitly so the 3 corresponding pNGBs are not well-protected by the shift symmetry. More discussion will be presented in the next subsection.} with pNGBs as the Higgs doublet.

The low energy theory for these bound states can be described by the Nambu-Jona-Lasinio (NJL) model \cite{Nambu:1961tp, Nambu:1961fr}. To construct the model, we start with the components that are involved in the strong interaction. For future convenience, we use $q'$ instead of $q$ to represent quarks in the flavor basis. Therefore, we have
\begin{equation}
\psi'_L=(T'_L,q'_L)^T=(T'_L,t'_L,b'_L)^T,\quad \psi'_R=(T'_R,t'_R)^T~,
\end{equation}
which participate in the strong interaction. The quark-anti-quark bound states $\Phi$ can be written in the forms of different groups as
\begin{equation}
\Phi=\bar{\psi'_R}\psi'_L=
\begin{pmatrix}
\bar{T'_R}\,T'_L & \bar{t'_R}\,T'_L \\
\bar{T'_R}\,q'_L & \bar{t'_R}\,q'_L\\
\end{pmatrix}
=
\begin{pmatrix}
\Phi_T & \Phi_t \\
\end{pmatrix}
=
\begin{pmatrix}
S_T & S_t \\
H_T & H_t \\
\end{pmatrix},
\end{equation}
where $\Phi_T$ and $\Phi_t$ are $SU(3)_L$ triplet, $S_T$ and $S_t$ are $SU(2)_L$ singlet and $H_T$ and $H_t$ are $SU(2)_L$ doublet.

\subsection{Scalar potential and vacuum structure}\label{section:VH}

For a realistic potential to realize the desired mechanism, we follow the construction in \cite{Cheng:2013qwa}. First, at the scale below $\Lambda$, the fermion coupling will generate the quadratic and quartic potential for the bound state $\Phi$ as
\begin{align}\label{VPhi}
V_{\Phi}=M_\Phi^2\,\Phi^\dagger\Phi+\frac{\lambda_1}{2}\text{Tr}[(\Phi^\dagger\Phi)^2]+\frac{\lambda_2}{2}(\text{Tr}[\Phi^\dagger\Phi])^2~.
\end{align}
Assuming the coupling of the strong interaction is greater than the critical coupling, then the coefficient of quadratic term $M_\Phi^2$ becomes negative, which can trigger the breaking of global symmetry $U(3)_L\times U(2)_R$.

To get the top seesaw mechanism, explicit $U(2)_R$ breaking terms are required, which can lead to the difference between $\Phi_T$ and $\Phi_t$. The effects can be parameterized as
\begin{align}\label{VR}
V_{\not{R}}=\delta M_{TT}^2\Phi_T^\dagger\Phi_T+\delta M_{tt}^2\Phi_t^\dagger\Phi_t+(M_{Tt}^2\Phi_T^\dagger\Phi_t+\text{ h.c.}).
\end{align}
Also, the tadpole terms for the singlet scalar are required, which can be generated from the quark mass term. Following the hindsight in \cite{Cheng:2013qwa}, we assume only the mass for $\bar{T'_L}\,t'_R$ is presented, which leads to a tadpole term for the corresponding bound state as
\begin{align}\label{Vtadpole}
\Delta\mathcal{L}=-m_{Tt}\bar{T'_L}\,t'_R+\text{ h.c.} ~ \implies ~
V_{\text{tadpole}}=-C_{Tt}S_t+\text{ h.c.}, ~\text{where}~C_{Tt}\propto{m_{Tt}}\,\Lambda^2~.
\end{align}
This term breaks not only the $U(2)_R$ but also the $U(3)_L$ explicitly which helps generate the desired vacuum misalignment.

The combined potential for the bound states is given by
\begin{align}\label{Vscalar}
V_{\text{scalar}}=V_{\Phi}+V_{\not{R}}+V_{\text{tadpole}}.
\end{align}
It introduces an overall global symmetry breaking scale $f$ with a specific vacuum structure
\begin{equation}\label{vacuum}
\langle\Phi\rangle=
\begin{pmatrix}
\langle S_T \rangle  & \langle S_t \rangle  \\ 
\langle H_T \rangle  & \langle H_t \rangle  \\
\end{pmatrix}=\frac{1}{\sqrt{2}}
\begin{pmatrix}
u_T  & u_t  \\ 
v  & 0  \\
 0 & 0 \\
\end{pmatrix},\quad \text{where}~f=\sqrt{u_T^2 + u_t^2  + v^2}~.
\end{equation}
Under this structure, we can see $\langle H_t \rangle$ is absent and $\langle H_T \rangle =v/\sqrt{2}$ is the only electroweak symmetry breaking source, which should then equal to $246$ GeV. Expanding the fields around their VEVs, we can write
\begin{align}
&S_T=\frac{1}{\sqrt{2}}\left(u_T+s_T+i\pi_T\right)~,
&S_t=\frac{1}{\sqrt{2}}\left(u_t+s_t+i\pi_t\right)\\
&H_T=
\begin{pmatrix}
\frac{1}{\sqrt{2}}\left(v+h_T+iA_T\right) \\ 
H_T^-\\ 
\end{pmatrix}~,
&H_t=
\begin{pmatrix}
\frac{1}{\sqrt{2}}\left(h_t+iA_t\right) \\ 
H_t^-\\ 
\end{pmatrix}
\end{align}
Since $\langle H_T \rangle$ is the only EWSB source, $A_T$ and $H_T^\pm$ are the longitudinal part of $Z$ and $W^\pm$ bosons. The $125$ GeV Higgs, on the other hand, is the mixture of four CP-even neutral scalars, $h_T$, $h_t$, $s_T$, and $s_t$. It is mostly given by $h_T$ and is light due to its pNGB nature. The approximate $U(3)_L\to U(2)_L$ symmetry breaking generates five pNGB in total. The last one is eaten by the additional $U(1)'$ gauge boson which will be discussed in section \ref{sec:U(1)'}.

\subsection{Top Yukawa coupling from the top seesaw mechanism}

One of the main challenges in composite Higgs models is generating a large top Yukawa coupling. In most of CHMs where the top sector and the Higgs and strong sector are separated, the top Yukawa coupling is generically suppressed due to its higher dimensional nature and some complicated mechanisms are required to reduce the suppression. On the other hand, if the top sector is already included in the strong sector and directly related to the formation of the Higgs doublet, the problem becomes the opposite. Generic top quark masses in this type of top condensation models \cite{Miransky:1988xi, Miransky:1989ds, Marciano:1989xd, Bardeen:1989ds} are too heavy ($\sim 600$ GeV) and thus are ruled out. The effort becomes looking for mechanisms to suppress the top mass to the observed value.

In our model, the suppression is realized by the top seesaw mechanism with the assistance of a new vector-like top quark $T$ \cite{Dobrescu:1997nm, Chivukula:1998wd, Cheng:2013qwa}. Starting from the NJL model, the fermions couple to the bound states with a large coupling as
\begin{equation}
\mathcal{L}_{\text{Yukawa}}=-y_T
\begin{pmatrix}
\bar{T'_L} & \bar{t'_L}  & \bar{b'_L}
\end{pmatrix}
\Phi
\begin{pmatrix}
{T'_R} \\ {t'_R}\\
\end{pmatrix},
\end{equation}
where the generic value of $y_T$ from Topcolor model \cite{Hill:1991at} is about $\sim 3.6$. In this study, we will set $y_T=3.6$ as our benchmark value.

With the vacuum structure determined in eq. \eqref{vacuum}, we can derive the mass matrix for the quark with charge $+2/3$, i.e. the top quark and the vector-like top quark. The matrix is given by
\begin{equation}
\mathcal{L}_{\text{mass}}=-\frac{y_T}{\sqrt{2}}
\begin{pmatrix}
\bar{T'_L} & \bar{t'_L}
\end{pmatrix}
\begin{pmatrix}
u_T  & u_t  \\ 
v  & 0  \\
\end{pmatrix}
\begin{pmatrix}
{T'_R} \\ {t'_R}\\
\end{pmatrix}
=
\begin{pmatrix}
\bar{T_L} & \bar{t_L}
\end{pmatrix}
\begin{pmatrix}
M_T  & 0 \\ 
0  & m_t  \\
\end{pmatrix}
\begin{pmatrix}
{T_R} \\ {t_R}\\
\end{pmatrix},
\end{equation}
where
\begin{equation}
M_T \approx \frac{1}{\sqrt{2}}y_Tf,\quad 
m_t \approx \frac{1}{\sqrt{2}}y_T\frac{u_tv}{u_T}~.
\end{equation}
The large $y_T$ in general should lead to a much heavier top quark but now it receives an additional suppression $u_t/u_T$. If they satisfy
\begin{equation}
\frac{u_t}{u_T}\approx \frac{y_t}{y_T}=\frac{1}{3.6}~,
\end{equation}
then we can reproduce the correct top quark mass $m_t \approx 173$ GeV.

Now the physical top quark $t$ is the mixture of the original top quark $t'$ and the vector-like top quark $T'$. The mixing between the flavor basis and the mass basis is characterized by the two angles $\theta_L$ and $\theta_R$ with
\begin{equation}
s_L\equiv\text{sin }\theta_L\approx v/u_T,\quad
s_R\equiv\text{sin }\theta_R\approx u_t/u_T~.
\end{equation}
Based on the hindsight from experimental constraints, two angles should both be small, which allows us to rewrite the VEVs as
\begin{equation}
u_T\approx f,\quad v\approx fs_L,\quad u_t\approx f s_R
\end{equation}
such that everything is in terms of the breaking scale $f$ and the two mixings $s_L$ and $s_R$.

\section{$Z'$ from the additional $U(1)_{L(-R)}$ gauge symmetry}\label{sec:U(1)'}

Besides the Higgs boson, the next lightest state in the spectrum is the fifth pNGB, which is mostly given by the $\pi_T$ field. If the corresponding $U(1)'$ symmetry is gauged, then it would be eaten, which will result in a massive $Z'$ boson. In general, $U(1)'$ can be any linear combination of $U(1)$ subgroup within the global symmetry $U(3)_L\times U(2)_R$ where the corresponding scalar $S_T$ is charged. The fermion component of $S_T$ is $\bar{T'_R}\,T'_L$ so the possible $U(1)$ subgroups include 
\begin{equation}
\mathbb{I}_{3\times3}, ~T_8=\frac{1}{2\sqrt{3}}
\begin{pmatrix}
-2  & 0 &  0  \\ 
0  & ~1~ &  0  \\ 
0  & 0 &  ~1~  \\ 
\end{pmatrix}\subset U(3)_L\quad \text{,and}\quad
\mathbb{I}_{2\times2}, ~\tau_3=\frac{1}{2}
\begin{pmatrix}
~1~  & 0   \\ 
0  & -1   \\ 
\end{pmatrix}\subset U(2)_R~.
\end{equation}
In this work, we consider two candidates for the new $U(1)'$ gauge symmetry.


First, we focus on the simplest choice - the $T_8$ of $U(3)_L$, which directly corresponds to the broken $U(1)'$ in the $SU(3)_L/SU(2)_L$ coset. To simplify the following analysis, we renormalize the generator to make the SM fermions carrying the unity charge $Q_{SM}=1$. The generator of the resulting $U(1)'$, which we call it $U(1)_L$, is given by
\begin{equation}\label{U(1)L}
U(1)_L:~
\begin{pmatrix}
-2  & 0 &  0  \\ 
0  & ~1~ &  0  \\ 
0  & 0 &  ~1~  \\ 
\end{pmatrix}~.
\end{equation}
The new interaction terms between the $Z'$ gauge boson and quarks are given by
\begin{align}\label{ZLintQ}
\mathcal{L}_{\text{int}}^{\text{quark}}=g_{Z'}Z'_\mu\,(&-2\bar{T'_L}\gamma^\mu T'_L +\,\bar{T'_L}\gamma^\mu t'_L + \bar{b'}_{L}\gamma^\mu b'_{L}\,)\\
=g_{Z'}Z'_\mu\,(&\,(-2c_L^2+s_L^2)\,\bar{T}_L\gamma^\mu T_L +
(-2s_L^2+c_L^2)\,\bar{t}_L\gamma^\mu t_L + \bar{b}_{L}\gamma^\mu b_{L}\nonumber\\
&+(-3\,c_Ls_L)\,\bar{T}_L\gamma^\mu t_L+(-3\,c_Ls_L)\,\bar{t}_L\gamma^\mu T_L)~.
\end{align}
We further assume the same extension and mechanism for the third generation lepton. A new vector-like neutrino $N$ is added, which is also charged under the $U(1)_L$ symmetry. The detail of the vector-like neutrino $N$ as well as the model for neutrino masses are beyond the scope of the present work and are left to a future study. In this work, we will treat the mass $M_N$ as a free parameter and focus on the $Z'$ interaction, which is given by
\begin{align}\label{ZLintL}
\mathcal{L}_{\text{int}}^{\text{lepton}}=g_{Z'}Z'_\mu\,(&\,(-2c_L^2+s_L^2)\,\bar{N}_L\gamma^\mu N_L +
(-2s_L^2+c_L^2)\,\bar{\nu_\tau}_L\gamma^\mu {\nu_\tau}_L + \bar{\tau}_{L}\gamma^\mu \tau_{L}\nonumber\\
&+(-3\,c_Ls_L)\,\bar{N}_L\gamma^\mu {\nu_\tau}_L+(-3\,c_Ls_L)\,\bar{\nu_\tau}_L\gamma^\mu N_L)~.
\end{align}
So far, we only consider the mixing between the third generation fermions and the vector-like fermions. There should also be mixings between them and the fermions of the first and second generations, which will lead to fruitful phenomenology in flavor physics. The details will be discussed in the section \ref{sec:Flavor}.

The mass of the $Z'$ boson is determined by the vacuum structure and the corresponding charge. Under $U(1)_L$, the charges of bound states are given by
\begin{align}\label{chargesL}
S_T,\, S_t:-2,~\quad H_T,\, H_t:1~.
\end{align}
The $Z'$ mass is then given by
\begin{equation}\label{ZLmass}
M_{Z'}=\sqrt{ \left(Q_{S_T}\,g_{Z'}u_T\right)^2 + \left(Q_{S_t}\,g_{Z'}u_t\right)^2+\left(Q_{H_T}\,g_{Z'}v\right)^2}
\approx \left(2\,g_{Z'}f\right)^2,
\end{equation}
where $Q_i$ is the charge of the scalar $i$. For future convenience, we define the $Z'$ scale
\begin{equation}\label{fL}
f'\equiv \frac{M_{Z'}}{g_{Z'}}\approx 2f~,
\end{equation}
which is relevant in the study of the $Z'$ phenomenology in the following sections. The relation is determined by the charge of the dominant vacuum, i.e. $\langle S_T\rangle$, which in this case carries charge $|Q_{S_T}|=2$.


Next, we consider the $U(1)'$ based on $L-R$, where the $L$ is the same as the $U(1)_L$ in eq. \eqref{U(1)L} and the $R$ comes from the generator
\begin{equation}\label{U(1)R}
U(1)_R:~
\begin{pmatrix}
-2 &  0  \\ 
0 &  ~2~  \\ 
\end{pmatrix}.
\end{equation}
Under the $U(1)_{L-R}$, the charge of each fermion becomes
\begin{align}\label{chargef}
T'_L:-2,~\quad q'_L=(t'_L,b'_L):1,~\quad T'_R:2,~\quad t'_R:-2~.
\end{align}
The reason for this setup is based on the issue in the scalar potential discuss in section \ref{section:VH}. The realistic potential requires the mass term $m_{Tt}\,\bar{T'_L}\,t'_R$ as shown in \eqref{Vtadpole}. However, the mass $m_{Tt}$ should originate from a new scale which might break the $U(1)'$ symmetry directly, unless the combination $\bar{T'_L}\,t'_R$ is neutral under the $U(1)'$, which requires $t'_R$ carrying charge $Q_{t'_R}=-2$. On the other hand, the combination $\bar{T'_R}\,T'_L$ must carry a nonzero charge to be broken, which requires $Q_{T'_R}\neq -2$. Therefore, the simplest choice is having an additional $U(1)_R$ as shown in eq. \eqref{U(1)R} from $\tau_3$ inside $U(2)_R$, under which $t'_R$ and $T'_R$ carry opposite charges so the both conditions are satisfied.

Under the new assumption, the interaction is now extended to the right-handed fermions as given by
\begin{align}\label{ZLRintQ}
\mathcal{L}_{\text{int}}^{\text{quark}}=g_{Z'}Z'_\mu\,(&\,-2\bar{T'_L}\gamma^\mu T'_L +\,\bar{T'_L}\gamma^\mu t'_L + \bar{b'}_{L}\gamma^\mu b'_{L}\,+2\bar{T'}_R\gamma^\mu T'_R -\,2\bar{t'}_R\gamma^\mu t'_R\,)\\
=g_{Z'}Z'_\mu\,(&\,(-2c_L^2+s_L^2)\,\bar{T'_L}\gamma^\mu T'_L +
(-2s_L^2+c_L^2)\,\bar{t}_L\gamma^\mu t_L + \bar{b}_{L}\gamma^\mu b_{L}\nonumber\\
&+(-3c_Ls_L)\,\bar{T}_L\gamma^\mu t_L+(-3c_Ls_L)\,\bar{t}_L\gamma^\mu T_L\nonumber\\
&+(2c_R^2-2s_R^2)\,\bar{T'}_R\gamma^\mu T'_R +
(2s_R^2-2c_R^2)\,\bar{t}_R\gamma^\mu t_R\nonumber\\
&+(-4c_Rs_R)\,\bar{T}_R\gamma^\mu t_R+(-4c_Rs_R)\,\bar{t}_R\gamma^\mu T_R)
\end{align}
The lepton sector is similar with simple transformation as
\begin{align}\label{ZLRintL}
\mathcal{L}_{\text{int}}^{\text{lepton}}=\mathcal{L}_{\text{int}}^{\text{quark}}
\,(\,T,\,t,\,b\, \to N,\,\nu_\tau,\,\tau\,)
\end{align}

The charges of the scalars under $U(1)_{L-R}$ are given by
\begin{align}\label{chargesLR}
S_T:-4,~\quad S_t:0,~\quad H_T:-1,~\quad H_t:3~.
\end{align}
We can again derive the mass of $Z'$ boson as
\begin{equation}\label{ZLRmass}
M_{Z'}=\sqrt{ \left(Q_{S_T}\,g_{Z'}u_T\right)^2 + \left(Q_{S_t}\,g_{Z'}u_t\right)^2+\left(Q_{H_T}\,g_{Z'}v\right)^2}
\approx \left(4\,g_{Z'}f\right)^2,
\end{equation}
and the $Z'$ scale of $U(1)_{L-R}$ as
\begin{equation}\label{fLR}
f'\equiv \frac{M_{Z'}}{g_{Z'}}\approx 4f~,
\end{equation}
where we get a different factor because now $|Q_{S_T}|=4$.

Notice that either case is not perfectly preserved in the complete scalar potential. $U(1)_L$ is also broken by $-C_{Tt}S_t$ tadpole term and $U(1)_{L-R}$ is broken by $M_{Tt}^2\Phi_T^\dagger\Phi_t$ term. The source of these violating terms will directly break the corresponding $U(1)'$ from another scale, which will introduce additional $Z'$ mass terms and affect our analysis. In this work, we assume the origin of these $U(1)'$ violating terms comes from the scale lower than the breaking scale $f$ such that the $M_{Z'}$ we derive is still valid. Under the assumption, we then have the $Z'$ boson and composite Higgs come from the same scale.


\section{$M_W$ and electroweak precision tests}\label{sec:EW}

An obvious shortage of $SU(3)/SU(2)$ CHMs is the lack of custodial symmetry protection, which will modify the measurement in the electroweak sector. On the other hand, it could provide an explanation for the deviation in the electroweak sector, too. Especially, the latest $W$ boson measurement by the CDF collaboration \cite{CDF:2022hxs} shows a $7\sigma$ discrepancy. These electroweak precision tests (EWPTs) are usually expressed in terms of $S$, $T$, and $U$ parameters~\cite{Peskin:1990zt, Peskin:1991sw}. The $U$ is usually fixed to zero because it is suppressed by an additional factor $M_{\rm new}^2/m_Z^2$, which is typically small for heavy new physics. The $W$ mass deviation is also interpreted in terms of these oblique parameters in many following analyses \cite{Lu:2022bgw, Strumia:2022qkt, deBlas:2022hdk, Bagnaschi:2022whn, Asadi:2022xiy}, which point out that a larger $T$ parameter is preferred. In this section, we will discuss the contribution to the $S$ and $T$ parameters from the new physics in our model.

\subsection{The $S$ parameter}
The leading contribution to the $S$ parameter comes from the mixing between the SM gauge bosons and the composite vector resonances $\rho$, which act as the gauge boson partners. The resulting value is estimated to be \cite{Agashe:2003zs,Agashe:2005dk,Giudice:2007fh}
\begin{align}
\Delta S \sim c_S~4\pi \frac{v^2}{M_\rho^2}  
\sim c_S~0.03 \left(\frac{5 \text{ TeV}}{M_\rho}\right)^{2},
\end{align}
where $c_S$ is an $\mathcal{O}(1)$ factor and $M_\rho$ is the mass of the vector resonances $\rho$.

There is also a contribution from the nonlinear Higgs dynamics due to the deviations of the Higgs couplings, which result in an incomplete cancellation of the electroweak loops~\cite{Barbieri:2007bh,Grojean:2013qca}. This contribution is given by
\begin{align}
\Delta S \sim \frac{1}{12\pi}\text{ ln}\left(\frac{M_\rho^2}{m_h^2}\right)~\frac{v^2}{f^2} 
\sim 0.05\text{ ln}\left(\frac{M_\rho}{m_h}\right)~\frac{v^2}{f^2}  ~,
\end{align}
which is proportional to $v^2/f^2$ and depends logarithmically on $M_\rho$.

\subsection{The $T$ parameter} 

The $T$ parameter parametrizes the amount of custodial $SU(2)$ breaking. In our model, The leading contribution to the $T$ parameter comes from the loop involving the vector-like top quark $T$, which is given by \cite{Chivukula:1998wd}
\begin{equation}
\Delta T=\frac{3s_L^2}{16\pi^2\alpha v^2}\left[s_L^2M_T^2+4c_L^2\frac{M_T^2m_t^2}{M_T^2-m_t^2}\text{ ln}\left(\frac{M_T}{m_t}\right)-(2-2s_L^2)m_t^2\right]~.
\end{equation}
Substituting the relation we have in our model, we can rewrite it as
\begin{equation}
\Delta T\approx\frac{3}{16\pi^2\alpha f^2}\left[\frac{1}{2}y_T^2v^2+4m_t^2\text{ ln}\left(\frac{y_Tf}{\sqrt{2}m_t}\right)-2m_t^2\right]~.
\end{equation}

Another important contribution comes from the $Z-Z'$ mixing, which will shift the $Z$ boson mass and lead to a positive contribution to the $T$ parameter as \cite{Allanach:2022bik}
\begin{align}
\Delta T \sim \frac{1}{\alpha}\left(\frac{4Q_H^2g_{Z'}^2}{g^2+g'^2}\frac{m_Z^2}{M_{Z'}^2}\right)
\sim \frac{1}{\alpha}\left(\frac{Q_{H_T}^2}{Q_{S_T}^2}\frac{v^2}{f^2}\right)~.
\end{align}
This contribution depends not only on the scale $f$ but also on the ratio $|Q_{H_T}/Q_{S_T}|$, which is $1/2$ for $U(1)_L$ and $1/4$ for $U(1)_{L-R}$. Therefore, a larger contribution is expected for $U(1)_L$ compared to $U(1)_{L-R}$, which will end up as a stronger bound on the scale $f$.

There is also a contribution from the modifications of the Higgs couplings to gauge bosons due to the nonlinear effects of the pNGB Higgs as \cite{Barbieri:2007bh,Grojean:2013qca}
\begin{align}
\Delta T \sim -\frac{3}{16\pi \text{ cos }\theta_W^2}\text{ ln}\left(\frac{M_\rho^2}{m_h^2}\right)~\frac{v^2}{f^2} 
\sim -0.16\text{ ln}\left(\frac{M_\rho}{m_h}\right)~\frac{v^2}{f^2}  ~,
\end{align}
which again depends on $v^2/f^2$ and is logarithmically sensitive to $M_\rho$. The contribution can partially cancel the other two positive contributions. Summing up all the contributions, we can then compare them with the experimental results Especially, we would like to find the required amount to explain the CDF $M_W$ measurement.

\subsection{The CDF $W$ mass and the constraint on the scale $f$}\label{sec:Wmass}

Many analyses on the oblique parameters based on the new $M_W$ measured by the CDF collaboration have been conducted \cite{Lu:2022bgw, Strumia:2022qkt, deBlas:2022hdk, Bagnaschi:2022whn, Asadi:2022xiy}. We follow the $S-T$ contour derived in \cite{Asadi:2022xiy}. The $T$ parameter within $2\sigma$ preferred region can ranged from $0.12\lesssim T \lesssim 0.42$ according to different $S$ from $S\sim 0$ to $S\sim 0.35$.

To determine the allowed parameter space for the scale $f$, we need to first fix the additional parameters in the equations of $\Delta S$ and $\Delta T$, which include the mass $M_\rho$ and the ratio $|Q_{H_T}/Q_{S_T}|$. The charge ratio is $1/2$ for $U(1)_L$ and $1/4$ for $U(1)_{L-R}$. On the other hand, the value of $M_\rho$ is related to the strong dynamics but the exact value is undetermined and thus is usually taken as a free parameter in the model. The $S$ parameter is mainly determined by the value of $M_\rho$, where a large $M_\rho$ points to the small $(S,T)$ region and a small $M_\rho$ points to the opposite end. In this work, we consider different $M_\rho$ such that the whole $2\sigma$ $S-T$ contour is roughly covered.

For the small $(S,T)$ region, we assume a heavy vector resonance with $M_\rho=10$ TeV. The $S$ parameter is $\lesssim 0.01$ so the bound on the scale $f$ only comes from the requirement that $T\gtrsim 0.12$. Considering different charge ratios for different $U(1)'$, we get an upper bound on the scale $f$ as
\begin{align}\label{upperbound}
f ~\lesssim~ 5.8 ~\text{TeV} \quad\text{for}~U(1)_L~,\qquad~
f ~\lesssim~ 4.6 ~\text{TeV} \quad\text{for}~U(1)_{L-R}~,
\end{align}
above which the $\Delta T$ is too small to explain the $W$ mass deviation.

On the other end, we take $M_\rho=3$ TeV and a large $c_S=4$ such that the $\Delta S\sim 0.35$. The bound on $f$ then come from the requirement that $T\lesssim 0.42$. Again we consider two $U(1)'$ cases, which give us the corresponding lower bounds on the scale $f$ as
\begin{align}\label{lowerbound}
f ~\gtrsim~ 3.0 ~\text{TeV} \quad\text{for}~U(1)_L~,\qquad~
f ~\gtrsim~ 2.4 ~\text{TeV} \quad\text{for}~U(1)_{L-R}~,
\end{align}
below which the resulting $M_W$ is too heavy.


\section{Flavor Phenomenology}\label{sec:Flavor}

The discussions in the previous section give us the bounds on the symmetry breaking scale $f$ from the electroweak precision tests. Next, we discuss the constraints from flavor physics, mainly due to the new $Z'$ boson. Because of its third-generation-philic nature, the relevant processes are largely from $B$ meson and $\tau$ lepton. Especially, it might provide an explanation for the observed neutral current B anomalies.

To simplify the analysis, we follow \cite{Chung:2021ekz, Chung:2021xhd} and assume all the nontrivial flavor violating processes originated from the left-handed sector. Under this assumption, there is no difference between the two $U(1)'$ because they share the same structure in the relevant left-handed part, which is given by the Lagrangian
\begin{align}\label{Zint0}
\mathcal{L}_{\text{int}}=g_{Z'}Z'_\mu\,(\,\bar{b}_L^f\gamma^\mu b_L^f+\bar{\tau}_L^f\gamma^\mu  \tau_L^f\,)~.
\end{align}
The $b_L^f$ and $\tau_L^f$ are still in the flavor basis and both relevant fields carry $Q=1$.

To study the flavor phenomenology of the $Z'$ boson, we need to extend the third-generation interaction to cover all the generations. Due to the observation of the nontrivial CKM matrix and PMNS matrix, we know the third-generation-philic interaction in the flavor basis can in general be transferred to the light generations by the rotation between flavor and mass basis.

In the flavor basis, the interaction terms are given by
\begin{align}\label{Zint1}
\mathcal{L}_{\text{int}}=g_{Z'}Z'_\mu\,(\,\bar{D}_L^f\gamma^\mu  Q_{D_{L}}^f D_L^f+\bar{E}_L^f\gamma^\mu  Q_{E_{L}}^fE_L^f\,)~,
\end{align}
where $D$ represents the three down-type quarks, $E$ represents the three charged leptons, and $f$ denotes the flavor basis. The charge matrices are given by
\begin{equation}\label{Zint1Charge}
Q_{D_{L}}^f=Q_{E_{L}}^f=
\begin{pmatrix}
0   &  0  &  0    \\
0   &  0  &  0    \\
0   &  0  &  1    \\
\end{pmatrix}.
\end{equation}

Next, we need the mixing matrices $U_{F_{L}}$ to transform them from the flavor basis $F^f_{L}$ to the mass basis $F^m_{L}$, which satisfy $F^f_{L}= U_{F_{L}} F^m_{L}$. The charge matrices in the mass eigenstate are then given by
\begin{equation}
Q_{F_{L}}^m=U_{F_{L}}^\dagger Q_{F_{L}}^fU_{F_{L}}.
\end{equation}
Following the analysis in \cite{Chung:2021ekz, Chung:2021xhd}, we only focus on the rotation matrices between the second and third generation of the down-type quarks, $U_{D_L}$, and charged leptons, $U_{E_L}$. The rotation matrices are given by
\begin{equation}
U_{D_L}=
\begin{pmatrix}
1   &0   &0    \\
0   &\text{cos}~\theta_d   &  \text{sin}~\theta_d \\
0   &-\text{sin}~\theta_d  &  \text{cos}~\theta_d \\
\end{pmatrix},\quad
U_{E_L}=
\begin{pmatrix}
1   &0   &0    \\
0   &\text{cos}~\theta_e   &  \text{sin}~\theta_e \\
0   &-\text{sin}~\theta_e  &  \text{cos}~\theta_e \\
\end{pmatrix}~.
\end{equation}

The resulting interaction terms are given in the mass basis as
\begin{align}
\mathcal{L}_{\text{int}}=g_{Z'}Z'_\mu\,(\,\bar{D}_L^m\gamma^\mu Q_{D_{L}}^mD_L^m+\bar{E}_L^m\gamma^\mu Q_{E_{L}}^mE_L^m\,)~,
\end{align}
where $D_{L}^m=(d_L,s_L,b_L)$, $E_{L}^m=(e_L,\mu_L,\tau_L)$, and the charge matrices are
\begin{equation}
Q_{D_{L}}^m=
\begin{pmatrix}
0   &  0  &  0    \\
0   &   \text{sin}^2\,\theta_d  &  -\, \text{sin}\,\theta_d\,\text{cos}\,\theta_d    \\
0   &  -\, \text{sin}\,\theta_d\,\text{cos}\,\theta_d  &   \text{cos}^2\,\theta_d    \\
\end{pmatrix},\quad
Q_{E_{L}}^m=
\begin{pmatrix}
0   &  0  &  0    \\
0   &   \text{sin}^2\,\theta_e  &  -\, \text{sin}\,\theta_e\,\text{cos}\,\theta_e    \\
0   &  -\, \text{sin}\,\theta_e\,\text{cos}\,\theta_e  &   \text{cos}^2\,\theta_e    \\
\end{pmatrix}~.
\end{equation}
Then, we can write down all the couplings with the $Z'$ boson for the left-handed fermions. To study the B anomalies, two of them, $Z'sb$ and $Z'{\mu\mu}$ are especially important so we define the corresponding couplings as $g_{sb}$ and $g_{\mu\mu}$. Moreover, we can extract the charge and mixing part of the couplings and rewrite the couplings as
\begin{align}
g_{sb}= -\,g_{Z'}\epsilon_{sb} &\quad\text{where}\quad
\epsilon_{sb}=\, \text{sin}\,\theta_d\,\text{cos}\,\theta_d~,\\
g_{\mu\mu}= g_{Z'}\epsilon_{\mu\mu}
&\quad\text{where}\quad\epsilon_{\mu\mu}=\text{sin}^2\,\theta_e~.
\end{align}
We will show later that the constraints from flavor physics can be put on the three key parameters: the $Z'$ scale $f'$, the mixings $\epsilon_{sb}$, and $\epsilon_{\mu\mu}$.

With these specified mixing matrices, we can then explore the consequence in flavor physics, especially the parameter space allowed to explain the neutral current B anomalies. Also, the constraints from other low energy experiments, including neutral meson mixings and lepton flavor violating decays, are discussed in this section.

\subsection{Neutral Current B Anomalies}

To explain the observed neutral current B anomalies, an additional contribution on the process $b\to s\mu\mu$ is required. Based on the assumption we made, after integrating out the heavy $Z'$ boson, we can get the operator
\begin{equation}
\Delta\mathcal{L}_{\text{NCBA}}=\frac{4G_F}{\sqrt{2}}V_{tb}V_{ts}^*\frac{e^2}{16\pi^2}\,C_{LL}(\bar{s}_L\gamma_\mu b_L)(\bar{\mu}_L\gamma^\mu\mu_L)
\end{equation}
in the low energy effective Lagrangian with coefficient
\begin{equation}
C_{LL}
=\frac{g_{sb}g_{\mu\mu}}{M_{Z'}^2}~(36~\text{TeV})^2
=-\frac{\epsilon_{sb}\epsilon_{\mu\mu}}{f'^2}~(36~\text{TeV})^2~.
\label{Banomaly}
\end{equation}

If we only consider the clean observables, the global fit value for the Wilson coefficient $C_{LL}$ is given by \cite{Altmannshofer:2021qrr}
\begin{equation}\label{fitting}
C_{LL}=-0.70\pm 0.16~,
\end{equation}
which requires
\begin{subequations}
\label{bsmumu}
\begin{equation}
\frac{\epsilon_{sb}\epsilon_{\mu\mu}}{f'^2}=\frac{1}{(43~\text{TeV})^2}
\implies f'= \sqrt{\epsilon_{sb}\epsilon_{\mu\mu}}~(43~\text{TeV})~.
\end{equation}
Besides, we also consider the value $C_{LL}=-0.38$, which is $2\sigma$ above the central value \footnote{Based on the latest measurement \cite{LHCb:2022zom}, the global fit value for the Wilson coefficient $C_{LL}$ considering only the clean observables becomes \cite{Greljo:2022jac}
\begin{equation}\label{fitting_new}
C_{LL}=-0.20\pm 0.14~,
\end{equation}
which is far from the original value and makes this analysis useless. The parameter space restricted by NCBAs is now available because now $C_{LL}$ is consistent with  zero. The originally-available parameter space is now shaded due to the lower bound on $C_{LL}>-0.48$ considering $2\sigma$ region.}. This value represents a smaller contribution from new physics but is of experimental and theoretical interest as we will see in the following sections. It requires
\begin{equation}
\frac{\epsilon_{sb}\epsilon_{\mu\mu}}{f'^2}=\frac{1}{(58~\text{TeV})^2}
\implies f'= \sqrt{\epsilon_{sb}\epsilon_{\mu\mu}}~(58~\text{TeV})~.
\end{equation}
\end{subequations}
The two different values of $C_{LL}$ will lead to very different phenomenology, especially in the direct search strategies, which will be discussed in the section \ref{sec:Collider}.

The generic scale for both cases shows the $Z'$ scale $f' \gtrsim 40$ TeV. But if the values of $\epsilon_{sb}$ and $\epsilon_{\mu\mu}$ are small, the exact scale will be much lower, which can match the scale preferred by the $W$ mass anomaly.

\subsection{Neutral Meson Mixing}

Next, we discuss the constraints from other measurements. The strongest constraints on the $Z'$ solution come from neutral meson mixings. In our specified mixing matrices, the mixings between the first generation down-type quark and other generations are not presented, so there are no additional contributions on $K^0-\bar{K}^0$ and $B_d-\bar{B}_d$ mixings. In the up-type quark sector, additional contribution on $D^0-\bar{D}^0$ mixing receives CKM suppression, so the constraint is also relaxed. Therefore,  the strongest bound on the $Z'$ coupling would come from the $B_s-\bar{B}_s$ mixing, which contains the second and third generation down-type quarks. Integrating out the heavy $Z'$, we get the operator
\begin{equation}
\Delta\mathcal{L}_{B_s}=-\frac{1}{2}\frac{g_{sb}^2}{M_{Z'}^2}(\bar{s}_L\gamma_\mu b_L)(\bar{s}_L\gamma_\mu b_L)~.
\end{equation}
Following the calculation in \cite{DiLuzio:2017fdq}, assuming $M_{Z'}$ is around the TeV scale, we can derive the deviation on the mass difference of neutral $B_s$ mesons as
\begin{equation}
C_{B_s}\equiv \frac{\Delta M_s}{\Delta M_s^{SM}}
\approx 1+\left(5576 \text{ TeV}^2\right)\,\left(\frac{g_{sb}}{M_{Z'}}\right)^2,
\end{equation}
The measurement of mixing parameter \cite{HFLAV:2019otj} compared with SM prediction by sum rule calculations \cite{King:2019lal} gives a strong upper bound at 95\% C.L. as \cite{Allanach:2019mfl}
\begin{equation}\label{Bsmixing}
\left|\frac{g_{sb}}{M_{Z'}}\right| \leq \frac{1}{194~\text{TeV}}\implies
\epsilon_{sb} \leq \frac{f'}{ 194~\text{TeV}}~.
\end{equation}

Next, we can combine it with the requirement from eq. \eqref{bsmumu}, which allows us to transfer the upper bound on $\epsilon_{sb}$ to the lower bound on $\epsilon_{\mu\mu}$ as
\begin{equation}\label{minmumu}
\epsilon_{\mu\mu} \geq \frac{f'}{ 9.4~(17.2)~\text{TeV}}~,
\end{equation}
where the value in the denominator corresponds to $C_{LL}=-0.70\,(-0.38)$ respectively. The transformation implies that, for the $b\to s\mu\mu$ process, the $bs$ side, which is constrained by the $B_s-\bar{B}_s$ mixing measurement, is extremely suppressed. Therefore, the $\mu\mu$ side needs to be large enough to generate the observed neutral current B anomalies. Therefore, with smaller $|C_{LL}|$, the required $ \epsilon_{\mu\mu}$ is also smaller.

Moreover, because there is a maximum value for $\epsilon_{\mu\mu}=$ sin$^2\,\theta_e\leq 1$, it implies an upper bound $9.4\,(17.2)$ TeV for the $Z'$ scale $f'$, which is consistent with the scale we expected. Also, we can combine eq. \eqref{Bsmixing} and eq. \eqref{minmumu}, which gives us the ratio $\epsilon_{\mu\mu}/\epsilon_{sb}\gtrsim 21\,(11)$. Again we put in $\epsilon_{\mu\mu}=$ sin$^2\,\theta_e\leq 1$, it then leads to the upper bound $\epsilon_{sb}\lesssim 0.05\,(0.09)$, which is consistent with the corresponding value we have in the CKM matrix.

\subsection{Lepton Flavor Violation}

\begin{figure}[t]
\centering
\includegraphics[width=1.0\linewidth]{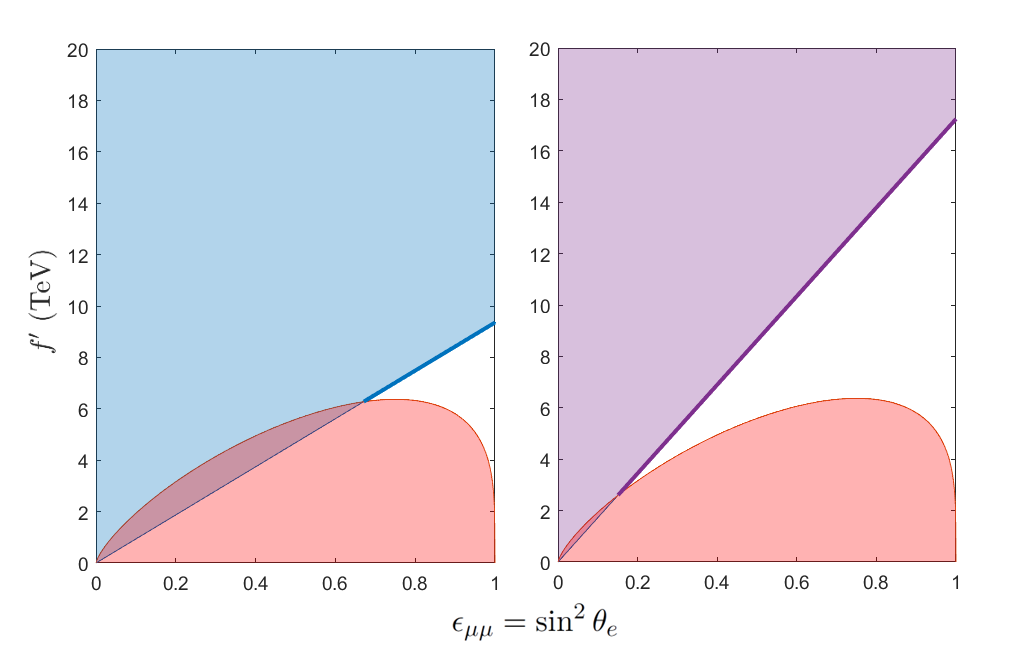}
\caption{The experimental constraints are shown in the plane of $f'$ vs. $\epsilon_{\mu\mu}$. The red shaded region is excluded by the constraints from the LFV decay $\tau\to 3\mu$. The blue (purple) shaded region in the left (right) is excluded by the constraint from $B_s-\bar{B}_s$ mixing combined with the requirements of NCBAs with $C_{LL}=-0.70\,(-0.38)$. The strong blue and purple line labels the boundary of the available parameter space in the two different cases, which are important when we discuss the direct searches in the dimuon channel.}
\label{LFV}
\end{figure}

In the lepton sector, there are also strong constraints from flavor changing neutral currents. The off-diagonal term in the lepton charge matrix $Q_{e_L}$ will introduce lepton flavor violating decays. Since we only specified the rotation between the second and the third generation, the most important effects show up in $\tau$ decays, especially the decay $\tau\to 3\mu$. The relevant term in the effective Lagrangian generated by integrating out the $Z'$ boson is given by
\begin{equation}
\Delta\mathcal{L}_{\text{LFV}}=\frac{g_{Z'}^2}{M_{Z'}^2}\text{sin}^3\,\theta_e \,\text{cos}\,\theta_e (\bar{\tau}_L\gamma^\rho\mu_L)(\bar{\mu}_L\gamma_\rho\mu_L)~,
\end{equation}
which leads to a branching ratio
\begin{align}
BR(\tau\to 3\mu)&=\frac{2m_\tau^5}{1536\pi^3\Gamma_\tau}\left(\frac{g_{Z'}^2}{M_{Z'}^2}\text{sin}^3\,\theta_e \,\text{cos}\,\theta_e\right)^2\nonumber\\
&=3.28\times10^{-4}\,\left(\frac{1\text{ TeV}}{f'}\right)^4\epsilon_{\mu\mu}^3(1-\epsilon_{\mu\mu})~.
\end{align}
The current bound on the branching ratio is $<2.1\times 10^{-8}$ at $90\%$ CL \cite{Hayasaka:2010np}, which requires
\begin{align}
\left(\frac{1\text{ TeV}}{f'}\right)^4\epsilon_{\mu\mu}^3(1-\epsilon_{\mu\mu})< 6.4\times10^{-5}~.
\end{align}
It puts a strong constraint on the available parameter space, especially around the middle value of $\epsilon_{\mu\mu}$. The exclusion plot combining the requirement of neutral current B anomalies and the constraint from $B_s-\bar{B}_s$ mixing in the plane of $f'$ vs. $\epsilon_{\mu\mu}$ is shown in figure~\ref{LFV}.

In the left plot, the white region is the allowed parameter space with $C_{LL}=-0.70$, where the small $\epsilon_{\mu\mu}$ region is totally excluded with a minimal value $\epsilon_{\mu\mu}\geq 0.67$. Therefore, we call it ``large angle region''. In the right plot with $C_{LL}=-0.38$, $\epsilon_{\mu\mu}\geq 0.15$ is allowed, which is called ``small angle region'' and ends up with a richer phenomenology.

Theoretically, the limit $\epsilon_{\mu\mu}=1$ implies that the muon should be the real third generation lepton in the flavor basis. Although we only have a little understanding about the origin of SM flavor structure, it still looks unnatural for such an assumption. From this point of view, the small $\epsilon_{\mu\mu}$ region is preferred and therefore worth more studies.

\begin{figure}[t]
\centering
\includegraphics[width=1.0\linewidth]{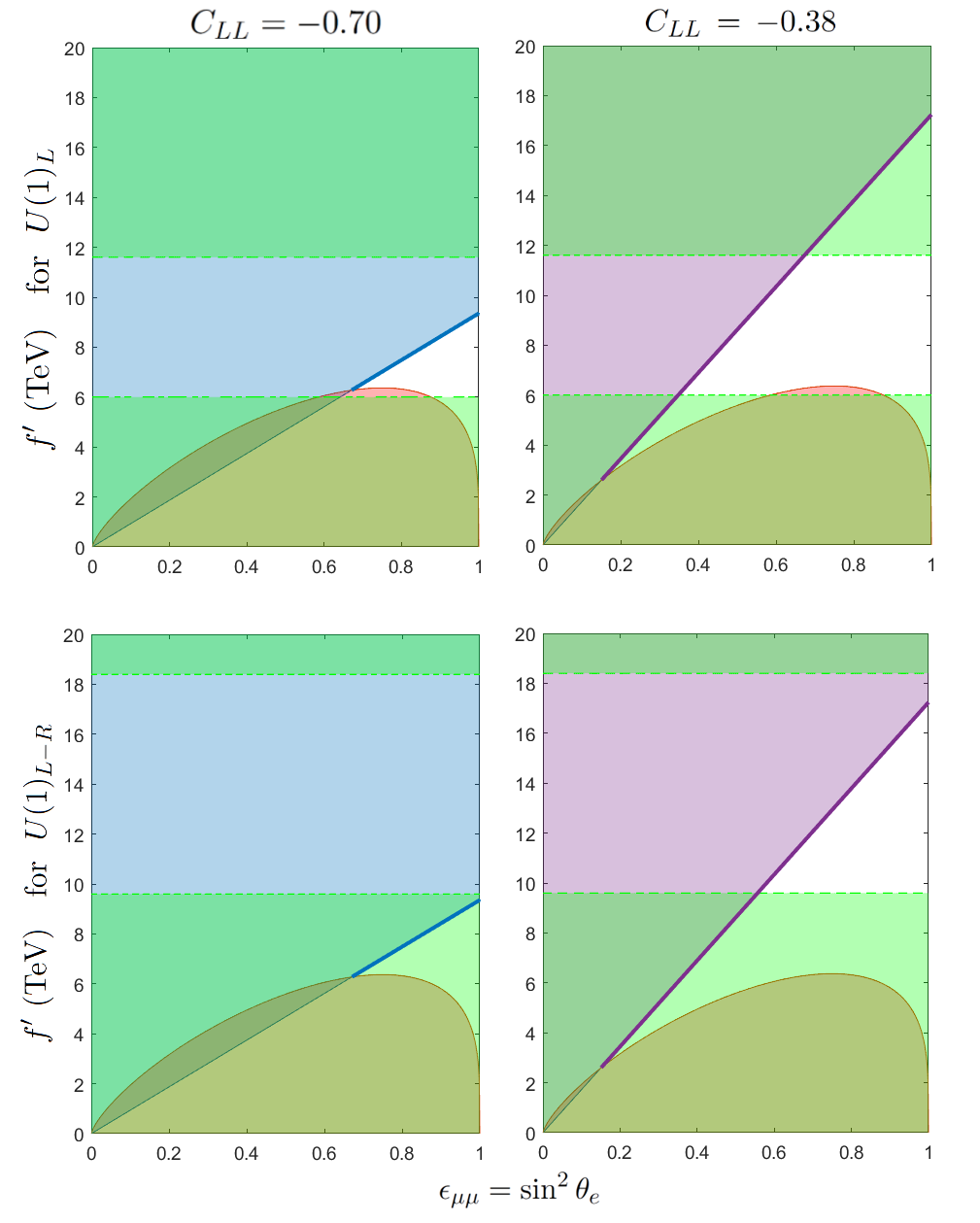}
\caption{The experimental constraints including the EWPTs and CDF $M_W$ measurement are shown in the plane of $f'$ vs. $\epsilon_{\mu\mu}$. Besides the constraints already shown in figure~\ref{LFV}. The bounds on $f'$ from eq. \eqref{boundL} and \eqref{boundLR} are also included as green shaded regions. The two plots on the top include the bounds for the model with gauged $U(1)_L$ and the two plots on the bottom are for the gauged $U(1)_{L-R}$ model. }
\label{LFV20_all}
\end{figure}


\section{Connecting the $B$ anomalies with the $M_W$ anomaly}\label{sec:Combine}

In section~\ref{sec:Wmass}, we found the preferred $f$ to explain the CDF $M_W$ for the two $U(1)'$ models. In the last section, the preferred $f'$ as a function of $\epsilon_{\mu\mu}$ are also presented. To figure out the possibility to solve both problems with the same UV physics, we combine the two conditions. Using the relation in eq. \eqref{fL} and eq. \eqref{fLR}, the bounds on the scale $f$ from section~\ref{sec:Wmass} can be transformed to the bounds on the $Z'$ scale $f'$. The bounds are given by
\begin{align}\label{boundL}
&6.0 ~\text{TeV}~\lesssim ~f' ~\lesssim~ 11.6 ~\text{TeV} \quad\text{for }~U(1)_L~,\\\label{boundLR}
&9.6 ~\text{TeV}~\lesssim ~f' ~\lesssim~ 18.4 ~\text{TeV} \quad\text{for }~U(1)_{L-R}~,
\end{align}
where we find a lower (higher) scale is preferred in the model with $U(1)_{L(-R)}$ symmetry. The combined plots are shown in figure~\ref{LFV20_all}.

For gauged $U(1)_L$, which prefers the lower scale, viable parameter spaces exist as a triangle region assuming the normal $|C_{LL}|$. A smaller $|C_{LL}|$ opens up a large parameter space, especially in the small $\epsilon_{\mu\mu}$ region, which will result in a richer phenomenology. On the other hand, the model with gauged $U(1)_{L-R}$, which prefers the higher scale, can not explain $W$ mass and the NCBAs at the same time. Some parameter spaces are allowed assuming a smaller $|C_{LL}|$ but still only in the parameter space with a large $\epsilon_{\mu\mu}$.


\section{Collider searches}\label{sec:Collider}

Several new degrees of freedom are introduced in our model, including the heavy Higgs doublet and new vector-like fermions. However, in this section, we only focus on the new $Z'$ boson, which should be the lightest among them. Also, the properties of the $Z'$ boson are partially fixed by the requirements to explain the two anomalies, which makes it predictable and a good target for direct searches.

The analyses of EWPTs and flavor physics from previous sections already put constraints on the mixings, the scale $f$ and $f'=M_{Z'}/g_{Z'}$. Direct searches, on the other hand, can provide the lower bound on the mass $M_{Z'}$ directly. A general $Z'$ collider search has been discussed in \cite{Allanach:2019mfl, Chung:2021xhd}, so in this section, we will focus on the parameter spaces available in this model, which are shown in the figure~\ref{LFV20_all}.

The direct searches depend on the $Z'$ properties. In this work, we consider two types of $U(1)'$ symmetry with different interaction terms, which lead to very different properties, especially the branching ratio. Also, the allowed parameter spaces are different. Therefore, we will discuss them separately.

Another undetermined issue is the neutrino sector. The properties of the vector-like neutrino $N$ and the right-handed neutrino $\nu_R$ will both affect the $Z'$ phenomenology. However, to determine them, a complete theory of neutrino mass is required, which is beyond the scope of this work. In the following analysis, we assume a Dirac-type neutrino such that $\nu_R$ is light and thus the decay $Z'\to \bar{\nu_R}\nu_R$ is allowed. On the other hand, for the vector-like neutrino $N$, we assume $M_N> M_{Z'}$ such that $Z'\to NN$ and $Z'\to N\nu$ are both forbidden. Therefore, both new vector-like fermions $T$ and $N$ are heavier than the $Z'$ boson so we can focus on the $Z'$ interaction with the SM fermions in the following discussion.

\subsection{The $Z'$ boson of $U(1)_L$ gauge symmetry}

Starting from the $U(1)_L$, the interaction between $Z'$ and the SM fermions is given by
\begin{align}
\mathcal{L}_{\text{int}}^{\text{SM}}
=\,g_{Z'}Z'_\mu\,((-2s_L^2+c_L^2)\,\bar{t}_L\gamma^\mu t_L + \bar{b}_{L}\gamma^\mu b_{L}+(-2s_L^2+c_L^2)\,\bar{\nu_\tau}_L\gamma^\mu {\nu_\tau}_L + \bar{\tau}_{L}\gamma^\mu \tau_{L})~.
\end{align}
Since the angle $\theta_L$ is small, it is approximately given by
\begin{align}
\mathcal{L}_{\text{int}}^{\text{SM}}
\approx g_{Z'}Z'_\mu\,(\,\bar{t}_L\gamma^\mu t_L + \bar{b}_{L}\gamma^\mu b_{L}+\,\bar{\nu_\tau}_L\gamma^\mu {\nu_\tau}_L + \bar{\tau}_{L}\gamma^\mu \tau_{L})~,
\end{align}
where the $Z'$ boson couples universally with all the thrid-generation SM left-handed fermions in the flavor basis. Using the specified matrices discussed in section \ref{sec:Flavor}, we can further transform the interaction to the mass basis.

The couplings to the first and second generation quarks are negligible due to the small mixing angles between flavor and mass basis. Therefore, the dominant production comes from the process $b_L\bar{b}_L$ fusion with the production cross section given by
\begin{equation}
{\sigma}(b_L\bar{b}_L\to Z') \equiv {g_{Z'}^2}\cdot \sigma_{b_Lb_L}(M_{Z'})~,
\end{equation}
where the coupling dependence is taken out. The $\sigma_{b_Lb_L}$ (as a function of $M_{Z'}$) is determined by the left-handed bottom-quark parton distribution functions \cite{Martin:2009iq, Alwall:2014hca}.

The partial width of the $Z'$ boson decaying into Weyl fermion pairs $\bar{f_i}f_j$ is given by
\begin{equation}
\Gamma_{ij}=\frac{C}{24\pi}g_{ij}^2M_{Z'}~,
\end{equation}
where $C$ counts the color degree of freedom and $g_{ij}$ is the coupling of $Z'\bar{f_i}f_j$ vertex. In the limit that all $m_f$ are negligible, we get the total relative width as
\begin{equation}
\frac{\Gamma_{Z'}}{M_{Z'}}=\frac{1}{3\pi}g_{Z'}^2\sim 11\,\%\cdot g_{Z'}^2~.
\end{equation}
The narrow width approximation is valid up to $g_{Z'}\sim 1.0$.

The dominant decay channels are the diquarks channel of the third generation with
\begin{equation}
Br(t\bar{t})\sim Br(b\bar{b}) \sim 37.5\%~.
\end{equation}
However, the main constraint is expected to be from dilepton channels which are rather clean. Based on the mixing matrices we specified, the branching ratios are given as
\begin{align}
&Br(\tau\tau) \sim 12.5~(1-\epsilon_{\mu\mu})^2~\%~, \\
Br(\ell^+\ell^-) \sim 12.5\% \quad \implies \quad
&Br(\mu\tau) \sim 25.0~\epsilon_{\mu\mu}(1-\epsilon_{\mu\mu})~\%~, \\
&Br(\mu\mu) \sim 12.5~\epsilon_{\mu\mu}^2~\%~.
\end{align}

From the production cross section and the branching ratios we got, we can calculate the cross section for the dilepton final state. For example, for the dimuon final state, the contour line for the cross section are straight lines in figure~\ref{LFV} with a fixed ratio $\epsilon_{\mu\mu}/f'$. The minimal value of $\epsilon_{\mu\mu}/f'$ that corresponds to the strong blue (purple) line in figure \ref{LFV} gives the smallest cross section for $b_L\bar{b}_L \to Z' \to \mu\mu$ process.\footnote{The dimuon cross section for the $Z'$ in this model is accidentally the same as the one in \cite{Chung:2021ekz, Chung:2021xhd} due to the same $\sigma \times Br$, which results in a similar bound. However, it is different for other dilepton channels.} The lower bound on $M_{Z'}$ can then be derived by comparing with the experimental results from the ATLAS \cite{ATLAS:2019erb} with an integrated luminosity of 139 fb$^{-1}$.\footnote{The bound by collider searches also weakly depends on the relative width ${\Gamma_{Z'}}/{M_{Z'}}$. We choose a narrow relative width $\sim 0.5\%$, which corresponds to $g_{Z'}\sim 0.6$. The value is roughly consistent with the ratio between the bound on the $M_{Z'}$ and the scale on the corresponding segment in figure \ref{LFV}.}

\begin{figure}[t]
\centering
\includegraphics[width=1.0\linewidth]{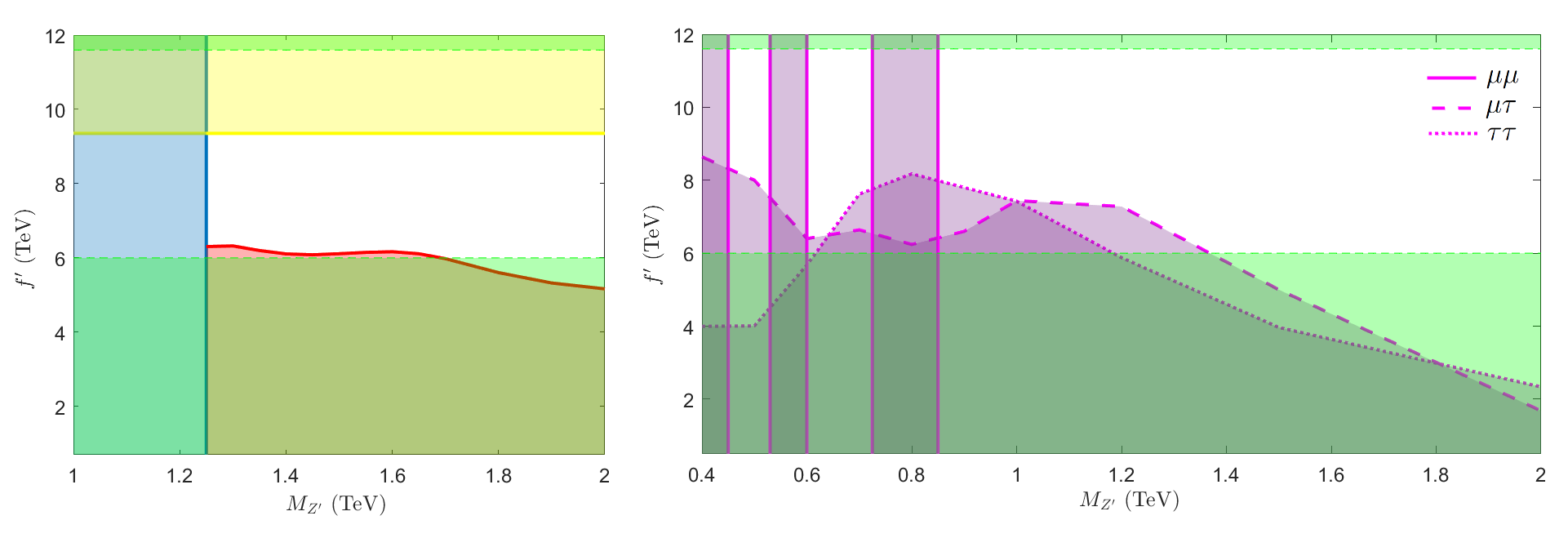}
\caption{Constraints in the $f'$ vs. $M_{Z'}$ plane for $M_{Z'}$ below $2$ TeV for $C_{LL}=-0.70$ (left) and $C_{LL}=-0.38$ (right). The white region is currently allowed, where $\epsilon_{\mu\mu}$ and $\epsilon_{sb}$ can be chosen to satisfy \eqref{Banomaly} from the requirement of the B anomalies. The yellow region on the left is the upper bound in the large angle region as derived in eq.~\eqref{minmumu}. The other shaded regions are excluded by the corresponding constraints in figure~\ref{LFV20_all} combined with the direct searches, where we use the ATLAS 139 fb$^{-1}$ $\mu\mu$ search \cite{ATLAS:2019erb}, CMS 137 fb$^{-1}$ $\mu\tau$ search \cite{CMS:2021tau}, and ATLAS 139 fb$^{-1}$ $\tau\tau$ search \cite{ATLAS:2020zms}.}. 
\label{fvsMzL}
\end{figure}

To analyze the full parameter space in the $f'$ v.s. $\epsilon_{\mu\mu}$ plane, we follow the method used in \cite{Chung:2021ekz, Chung:2021xhd}, which allows us to transform it to the constraints on $f'$ v.s. $M_{Z'}$ plane - the two most important parameters for our $Z'$ boson. The resulting plots on two different $C_{LL}$ values are shown in figure~\ref{fvsMzL}.

For $C_{LL}=-0.70$, since we already showed $\epsilon_{\mu\mu}\geq 0.67$ from the flavor constraints in figure \ref{LFV}, the $\mu\mu$ final state is all we need to consider. The current bound by the $\mu\mu$ search of ATLAS \cite{ATLAS:2019erb} with an integrated luminosity of 139 fb$^{-1}$ is given by $M_{Z'}\gtrsim 1250$ GeV, which is shown as the blue line in the figure (left). The constraints on the scale $f'$ come from (1) Yellow: the requirement to explain the NCBAs combined with the $B_s-\bar{B}_s$ meson mixing (2) Red: $\tau\to 3\mu$ LFV decays (3) Green: the requirements to explain the CDF $M_W$ measurement. A rough rectangle region is left as the viable parameter space.

For the small angle region with $C_{LL}=-0.38$, which has $\sim 3.5$ times smaller dimuon cross section, the $M_{Z'}$ could be as low as $400$ GeV, which is shown as the purple shaded region with a solid line in the figure (right). However, in this parameter space, the $\mu\tau$ and $\tau\tau$ channels become important and could also be the discovery channels. Comparing the cross section for each channel derived from the parameter space with the experimental results, $\mu\tau$ search taken from the CMS \cite{CMS:2021tau} with an integrated luminosity of 137 fb$^{-1}$ and $\tau\tau$ search taken from the ATLAS \cite{ATLAS:2020zms} with an integrated luminosity of 139 fb$^{-1}$, we get the purple shaded region with a dashed line and a dotted line excluded by $\mu\tau$ search and $\tau\tau$ search respectively. The overall constraints are much weaker and now a sub-TeV $Z'$ is also viable in the gap between the purple shaded regions.

Other decay channels are unlikely to be the discovery channel. However, if the $Z'$ boson is found, looking for the same resonance in other channels will be the priority. Through the searches in different final states, we can decide the couplings of the $Z'$ boson to other SM fields. The structure of couplings can help us distinguish between different $Z'$ models. In this model, the $Z'$ boson couples universally to all the third generation SM left-handed fermions in the flavor basis. After the transformation to the mass basis, the universality no longer exists, but a unique partial width ratio will be preserved as
\begin{equation}
\Gamma_{tt}:\Gamma_{bb}:\Gamma_{\ell\ell}\sim 3:3:1,
\end{equation}
where $\Gamma_{\ell\ell}$ is the sum of all the charged lepton partial widths. This ratio is an important prediction of our $Z'$ model.\footnote{The ratio happens to be the same as the ratio in the $Z'$ model of \cite{Chung:2021ekz, Chung:2021xhd}, where the $Z'$ boson has a vector-like universal coupling with all the thrid-generation SM fermions in the flavor basis. One can still distinguish the two models by checking the exact value of cross section or differential cross section over the angular distribution.} Moreover, since an anomaly-free $U(1)'$ within SM fermions can not achieve this ratio, it can be indirect evidence for the existence of additional fermions beyond the SM, such as the vector-like fermions in our model.


\subsection{The $Z'$ boson of $U(1)_{L-R}$ gauge symmetry}

The interaction between the $Z'$ of $U(1)_{L-R}$ symmetry and the SM fermions, assuming both angles $\theta_L$ and $\theta_R$ are negligible, is approximately given by
\begin{align}
\mathcal{L}_{\text{int}}^{\text{SM}}
\approx g_{Z'}Z'_\mu\,(\,\bar{t}_L\gamma^\mu t_L + \bar{b}_{L}\gamma^\mu b_{L}-2\,\bar{t}_R\gamma^\mu t_R +\,\bar{\nu_\tau}_L\gamma^\mu {\nu_\tau}_L + \bar{\tau}_{L}\gamma^\mu \tau_{L}-\,2\,\bar{\nu_\tau}_R\gamma^\mu {\nu_\tau}_R\,)~.
\end{align}
There are additional couplings with right-handed top quark and $\tau$ neutrino compared to the $U(1)_{L}$ case, which will lead to a different $Z'$ phenomenology.

The dominant production process is the same as the $Z'$ boson from $U(1)_{L}$ symmetry, which comes from the process $b\bar{b}\to Z'$. The cross section is also given by
\begin{equation}
{\sigma}(b_L\bar{b}_L\to Z') \equiv {g_{Z'}^2}\cdot \sigma_{b_Lb_L}(M_{Z'})~.
\end{equation}
The total relative width in the limit that all $m_f$ are negligible is 
\begin{equation}
\frac{\Gamma_{Z'}}{M_{Z'}}=\frac{1}{\pi}g_{Z'}^2\sim 32\,\%\cdot g_{Z'}^2~,
\end{equation}
where the narrow width approximation is only valid up to $g_{Z'}\sim 0.6$. The value is smaller than the $U(1)_{L}$ case due to the additional decay channels raised from the coupling with the right-handed fermions.

The dominant decay channel now becomes $t\bar{t}$ only. The three largest branching ratios are given as follows
\begin{equation}
Br(t\bar{t})\sim 62.5\%~ \quad Br(\nu\bar{\nu}) \sim 20.8\%~ \quad Br(b\bar{b}) \sim 12.5\%~.
\end{equation}
However, the main constraint is still expected to come from the clear dilepton channels, whose branching ratios are given by
\begin{align}
&Br(\tau\tau) \sim 4.16~(1-\epsilon_{\mu\mu})^2~\%~, \\
Br(\ell^+\ell^-) \sim 4.16\% \quad \implies \quad
&Br(\mu\tau) \sim 8.32~\epsilon_{\mu\mu}(1-\epsilon_{\mu\mu})~\%~, \\
&Br(\mu\mu) \sim 4.16~\epsilon_{\mu\mu}^2~\%~.
\end{align}
Based on the production and branching ratio we derive, we can again calculate the cross section for each dilepton channel.

\begin{figure}[t]
\centering
\includegraphics[width=0.7\linewidth]{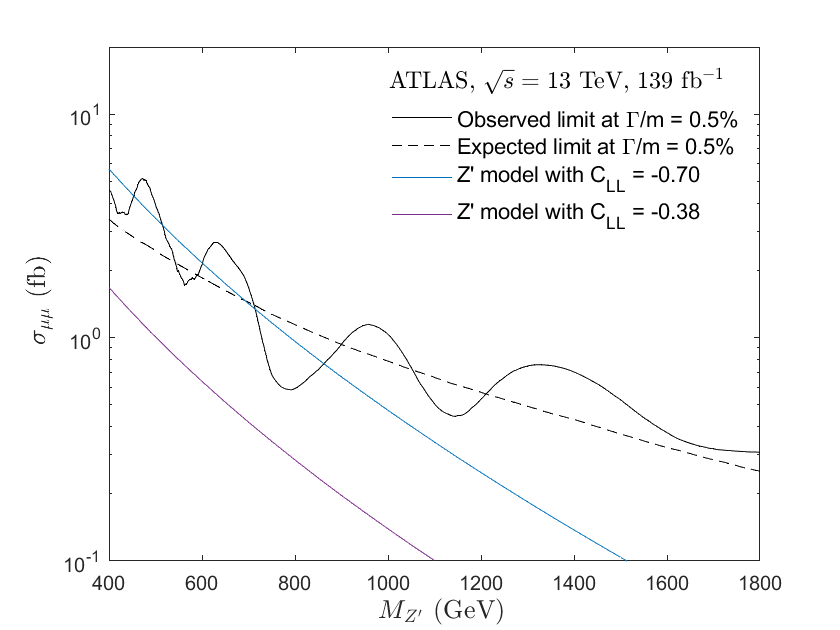}
\caption{Upper limits at 95\% CL on the cross section times branching ratio $\sigma_{\mu\mu}$ as a function of $M_{Z'}$ for 0.5\% relative width signals for the dimuon channel (black). Observed limits are shown as a solid line and expected limits as a dashed line. Also shown are theoretical predictions of the minimal cross section for $Z'$ boson from $U(1)_{L-R}$ gauge symmetry assuming $C_{LL}=-0.70$ (blue)  and $-0.38$ (purple).}
\label{mumuLR}
\end{figure}

For the large $|C_{LL}|$, there is no parameter space left because the preferred scale from CDF $M_W$ is too high to generate enough $C_{LL}$ to explain the neutral current B anomalies under the constraints of $B_s-\bar{B}_s$ meson mixing. Therefore, it is only possible considering a smaller $|C_{LL}|$. Still, due to the higher preferred scale, the allowed $\epsilon_{\mu\mu}$ is larger. Therefore, we only need to focus on the $\mu\mu$ final state. We calculate the dimuon cross section for two $C_{LL}$ and compare them with the direct search result from ATLAS \cite{ATLAS:2019erb} with an integrated luminosity of 139 fb$^{-1}$. The result is shown in figure \ref{mumuLR}. The constraint on the $M_{Z'}$ is much weaker than the bound we got in the $U(1)_L$ case due to the smaller $Br(\mu\mu)$. It turns out there are no constraints at all for heavy $Z'$ with a mass over $400$ GeV, which can only be explored during the HL-LHC era \cite{ATL-PHYS-PUB-2018-044, CidVidal:2018eel}.

Other decay channels, especially the $t\bar{t}$ final state, can be as large as $~50$ times $\mu\mu$ final state. However, even with such a large cross section, the current bound on the $Z'$ boson from the $t\bar{t}$ direct search \cite{ATLAS:2020lks} is still weaker than the bound from $\mu\mu$ direct search. Still, it is important to measure the ratio between different channels once the $Z'$ boson is discovered. For the $Z'$ boson of $U(1)_{L-R}$, the partial width ratio becomes
\begin{equation}
\Gamma_{tt}:\Gamma_{bb}:\Gamma_{\ell\ell}\sim 15:3:1,
\end{equation}
where $\Gamma_{\ell\ell}$ is the sum of all the charged lepton partial widths. This ratio is quite different from the one in the $U(1)_L$ case due to the additional coupling with $t_R$.


\section{Conclusions}\label{sec:Conclusion}

The naturalness principle so far is still the strongest motivation for particle physicists to look for TeV-scale new physics. Although lack of evidence from any direct searches, two anomalies appear from the precision measurement and both point to new degrees of freedom at the TeV scale. One is the long-standing neutral current B anomalies, which can trace back to almost ten years ago. The other is the CDF $M_W$ measurement, which is presented early this year with large significance. A desired UV model at the TeV scale should be able to address all three problems together.

To explain all of them, we consider a $U(3)_L/U(2)_L$ Top Seesaw model with an additional $U(1)'$ gauge symmetry. The Higgs doublet arises as pNGBs of the global symmetry breaking $U(3)_L\to U(2)_L$, which is naturally lighter than the breaking scale $f\sim$ few TeVs. Moreover, the top seesaw mechanism can naturally generate the top Yukawa coupling with $y_t\sim1$. The fifth pNGB is eaten by the $U(1)'$ gauge boson, which results in a heavy $Z'$ boson. Two possible candidates for $U(1)'$ symmetry, $U(1)_{L}$ and $U(1)_{L-R}$, are discussed in this work. This third-generation-philic $Z'$ boson can provide an explanation for the neutral current B anomalies. The lack of custodial symmetry, which used to be the fault of the model, generically predicts a heavier $M_W$ as observed by the CDF collaboration.

The viable parameter spaces preferred by the two anomalies are discussed in detail. The breaking scale $f$ between 3.0 (2.4) TeV and 5.8 (4.6) TeV in the $U(1)_{L(-R)}$ model can provide an explanation for the CDF $M_W$ measurement. Combined with the requirement from NCBAs and the two main constraints from FCNCs, $B_s-\bar{B}_s$ mixing and $\tau\to 3\mu$, which put the bounds on the $Z'$ scale $f'$, the mixing $\epsilon_{sb}$ and $\epsilon_{\mu\mu}$, we find that only the $Z'$ boson from the $U(1)_L$ gauge symmetry can address both anomalies at the same time. On the other hand, the $U(1)_{L-R}$ model can only address the conditions either with a smaller $|C_{LL}|$ or an even heavier $M_W$.

Direct searches of the new $Z'$ boson are also studied. The bound on the $M_Z'$ for the $U(1)_L$ model is $1.25$ TeV. A sub-TeV $Z'$ boson is also possible if $|C_{LL}|$ is $2\sigma$ lower. For the $U(1)_{L-R}$ case, considering the smaller $|C_{LL}|$, there is no bound from the direct searches due to its small branching ratio in the dimuon channel. The most important message from the bounds is: we are just about to explore the most relevant parameter space in this type of $Z'$ model. Different from the flavor universal new physics, which is almost ruled out up to several TeV by the first two runs of LHC, the third-generation-philic particles are just about to be tested in Run 3 and the upcoming HL-LHC. Together with updates on NCBAs and $M_W$, the mystery of the TeV-scale UV-complete model might be unveiled.

\acknowledgments

I thank Hsin-Chia Cheng and Florian Goertz for many useful discussions. I am also grateful to Ben Allanach and Wolfgang Altmannshofer for reading the previous paper and giving many helpful suggestions. I would like to express special thanks to the Mainz Institute for Theoretical Physics (MITP) of the Cluster of Excellence PRISMA+ (Project ID 39083149), for its hospitality and support.

\bibliographystyle{jhepbst}
\bibliography{Top_Seesaw_Ref}{}

\end{document}